\documentclass[useAMS,usenatbib]{mn2e}
\usepackage{amssymb}
\usepackage{graphicx}
\usepackage{lscape}
\usepackage{longtable}
\usepackage{array}
\title[The IGIMF in early- and late-type galaxies]{The galaxy-wide IMF of dwarf late-type to massive early-type galaxies}
\author[C.~Weidner et al.]
{C.~Weidner$^{1,2,3}$\thanks{E-mail: cweidner@iac.es}, P.~Kroupa$^{4}$\thanks{E-mail: pavel@astro.uni-bonn.de}, J.~Pflamm-Altenburg$^{4}$\thanks{E-mail: jpflamm@astro.uni-bonn.de} and A.~Vazdekis$^{1,2}$\thanks{E-mail: vazdekis@iac.es}\\
$^{1}$Instituto de Astrof{\'i}sica de Canarias, Calle V{\'i}a L{\'a}ctea s/n, E38205, La Laguna, Tenerife,
Spain\\
$^{2}$Dept. Astrof{\'i}sica, Universidad de La Laguna (ULL), E-38206 La Laguna, Tenerife, Spain\\
$^{3}$Scottish Universities Physics Alliance (SUPA), School of Physics and
  Astronomy, University of St. Andrews, North Haugh,\\
St. Andrews, Fife KY16 9SS, UK\\
$^{4}$Helmholtz-Institut f{\"u}r Strahlen- und Kernphysik (HISKP), Universit{\"a}t Bonn, Nussallee 14-16, D-53115 Bonn, Germany
}

\begin{document}
\bibliographystyle{aa}
\date{Accepted . Received 2013; in original form }

\pagerange{\pageref{firstpage}--\pageref{lastpage}} \pubyear{2013}

\maketitle

\label{firstpage}

\begin{abstract}
Observational studies are showing that the galaxy-wide stellar initial mass function are top-heavy in galaxies with high star-formation rates (SFRs). Calculating the integrated galactic stellar initial mass function (IGIMF) as a function of the SFR of a galaxy, it follows that galaxies which have or which formed with SFRs $>$ 10 $M_\odot$ yr$^{-1}$ would have a top-heavy IGIMF in excellent consistency with the observations. Consequently and in agreement with observations, elliptical galaxies would have higher M/L ratios as a result of the overabundance of stellar remnants compared to a stellar population that formed with an invariant canonical stellar initial mass function (IMF). For the Milky Way, the IGIMF yields very good agreement with the disk- and the bulge-IMF determinations. Our conclusions are that purely stochastic descriptions of star formation on the scales of a pc and above are falsified. Instead, star formation follows the laws, stated here as axioms, which define the IGIMF theory. We also find evidence that the power-law index $\beta$ of the embedded cluster mass function decreases with increasing SFR. We propose further tests of the IGIMF theory through counting massive stars in dwarf galaxies.
\end{abstract}



\section{Introduction}
\label{se:intro}
The stellar initial mass function (IMF), $\xi(m)$ = d$N$/d$m$, describes the distribution of masses of stars, whereby d$N$ is the number of stars formed in the mass interval [$m$, $m + \mathrm{d}m$]. It is one of the most important distribution functions in astrophysics as stellar evolution is mostly determined by stellar mass. The IMF therefore regulates the chemical enrichment history of galaxies, as well as their mass-to-light ratios and influences their dynamical evolution. Theoretically unexpected, the IMF is found to be invariant through a large range of conditions like gas densities and metallicities \citep{Kr01,Kr02,Ch03,EKW08,BCM10,MKD10,KWP13} and is well described by the canonical IMF (Appendix~\ref{app:IMF}). Therefore an invariant IMF is widely used to not only describe individual star clusters but also the stellar populations of whole galaxies. But, the question remains whether the IMF, derived from and tested on star cluster scales, is the appropriate stellar distribution function for complex stellar populations like galaxies. 

Several recent observations \citep{HG07,MWK09,LGT09,GHS10,DKB09,DKP12,CMA12,FBR13} have cast doubt on the existence of a universal IMF in galaxies. 

With this contribution we attempt to provide a single unifying principle for understanding the observationally deduced variation of the IMF from dwarf irregular to massive elliptical galaxies assuming the stellar IMF to be largely form-invariant in the star-forming building blocks of galaxies. These building blocks are the molecular cloud density peaks that form from a few to many millions of stars within $\approx$ 1 Myr and within $\approx$ 1 pc.

The observed variations of the galaxy-wide IMF in late-type and early-type galaxies are addressed in Section~\ref{se:obs} and these observational results are explained in the framework of the IGIMF effective theory in Section~\ref{se:model}. Here also the validity of the IGIMF for the Milky Way disc and Bulge are discussed. Section~\ref{se:res} presents the expected number of O and B stars for whole galaxies in dependence of the SFR for the IGIMF including star-bursting systems, and for a constant canonical IMF. Finally, the results are discussed in Section~\ref{se:diss}.\\

\section{Observational evidence for a varying galaxy-wide IMF in galaxies}

While for a long time the IMF of galaxies was assumed to be constant and observations didn't find evidence for any striking differences, it has been established over the last few years that the issue of IMF variation is more complex than previously thought. This has been made possible by the availability of very large and deep surveys, like the SDSS, which provide large populations of galaxies, allowing a statistical analysis of the effect of the IMF on galaxy properties.

\subsection{Observational evidence for a varying galaxy-wide IMF in late-type galaxies}
\label{se:obs}

\begin{figure}
\begin{center}
\includegraphics[width=9cm]{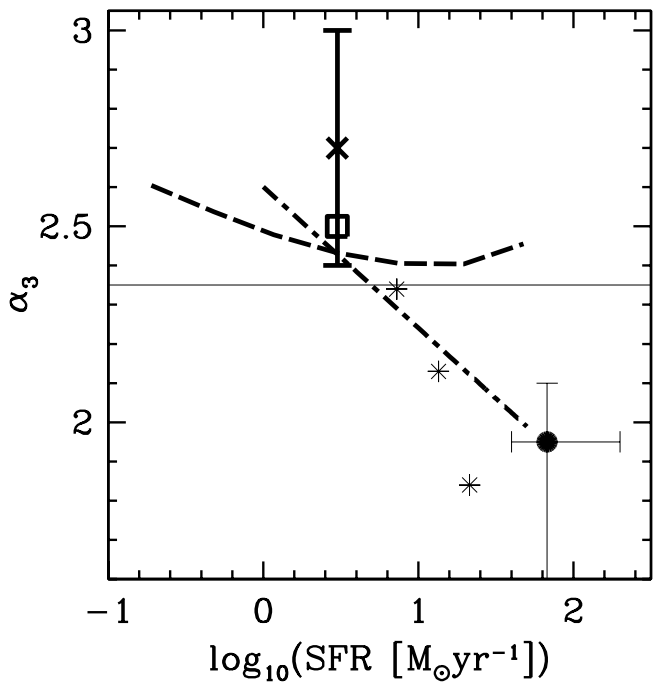}
\vspace*{-1.5cm}
\caption{The dependence of the slope of the observationally deduced galaxy-wide IMF above 1 $M_\odot$, $\alpha_3$, on the SFR. The dashed line is derived from the \citet{HG07} SDDS $r$-band magnitudes when using a fixed M/L ratio and a population build-up time of 12 Gyr. The long-dash--short-dashed line is the fit for late-type galaxies as deduced by the GAMA-team \citep{GHS10}. A large cross marks the \citet{Sc86} result for the Milky Way field derived from the present-day mass function \citep{KTG93} and the large box is the \citet{Ke83} value for the Milky Way. The three star symbols are results from \citet{Da07} while the big dot is from \citet{BKM07}. The thin horizontal line marks the Salpeter/Massey slope, $\alpha$ = 2.35 \citep{Sal55,Mass03}.}
\label{fig:GAMA}
\end{center}
\end{figure}

Using the large amount of data provided by the SDSS survey, \citet{HG07} derived IMF slopes above 1~$M_\odot$ ($\alpha_3$; $\alpha_3$ = 2.3 for the Salpeter value) for thousands of star-forming galaxies in dependence of their $r$-band magnitudes, $M_r$. They found that fainter galaxies have steeper IMF slopes than brighter (more massive) ones. In order to translate this slope dependence on $M_r$ into a dependence on galaxy mass, a fixed mass-to-light (M/L) ratio of 2 \citep{BMK03} was used together with a population build-up time of 12 Gyr. The dashed line in Fig.~\ref{fig:GAMA} shows the resulting behaviour for the SFRs derived when distributing the mass of the galaxy over the assumed age of the population of 12 Gyr.

Further evidence for IMF variations where found in 2009 when two teams of researchers \citep{MWK09,LGT09} discovered independently of each other that the H$\alpha$ fluxes of galaxies with very low star-formation rates do not match their expected levels when extrapolating from the UV fluxes. The ratio between H$\alpha$ to UV flux in these galaxies correlate with the galaxy-wide SFRs. Both groups reached the same conclusion that the IMF of these galaxies must vary systematically with physical parameters such as the SFR.

The {\it GA}laxy {\it M}ass {\it A}ssembly team \citep{DNB09,RDN10,BRH10} recently studied a large sample of nearly 44000 late-type galaxies with multi-colour photometry and spectroscopic redshifts. While attempting to model the stellar populations of these galaxies, \citet{GHS10} found that a flatter than Salpeter slope was necessary to do so and that this slope becomes increasingly flatter with increasing SFR. \citet{GHS10} used the derived slopes and the measured SFRs to directly fit a relation for the dependence of IMF slope above 1 $M_\odot$, $\alpha_3$, on the SFR. This relation is plotted as a long-dash--short-dashed line in Fig.~\ref{fig:GAMA}. Additionally shown in the figure are a number of different galaxy-wide IMF estimations. The cross with error bars is the Scalo-IMF derived from the field present-day Milky-Way (MW) mass function \citep{Sc86}, while the box is the IMF adopted by \citet{Ke83} to explain the H$\alpha$ flux of a number of star-forming galaxies. The dot with error bars is the \citet{BKM07} IMF determination from detailed chemical evolution modelling of the MW and M31 bulges (see \S~\ref{se:MW} below). They find a best-fitting model with $\alpha_3$ = 1.95 and find that the slope could be at most $\alpha_3$ = 2.1. The lower limit for $\alpha_3$ is less constraint as the overall metallicity can be reproduced with an IMF as flat as $\alpha_3$ = 1.33 but the authors also state that this would lead to some amount of oxygen overproduction. Furthermore, plotted as three star symbols is the \citet{Da07} result which is derived from comparing hydrosimulations of star-forming galaxies with observations to explain the evolution of the stellar-mass--SFR relation from redshift $z$ $\approx$ 2 to $z$ = 0. \citet{Da07} assumes a redshift dependent characteristic mass, $m^\ast$, of the IMF (instead of a changing mass function) of the form $m^\ast = 0.5 (1+z)^2$. Below $m^\ast$ the slope is set to 1.3 and above to 2.35. This can be transformed into a high-mass slope dependence with redshift by calculating the mean mass of an IMF with changing slope above a fixed $m^\ast$ of 0.5 $M_\odot$ and comparing this mean mass with the one obtained when using the redshift depended $m^\ast$. When both mean masses agree for a given redshift, the slope for the fixed $m^\ast$ is assigned to this redshift. \citet{Da07} defines a star-formation activity parameter, $\alpha_\mathrm{sf} = (M_\ast/SFR) / (t_\mathrm{H} - 1 \mathrm{Gyr})$ with $M_\ast$ = 10$^{10.7}$ $M_\odot$ and $t_\mathrm{H}$ = 13.1698 Gyr. The author describes this parameter as {\it the fraction of the Hubble time (minus a Gyr) that a galaxy needs to have formed stars at its current rate in order to produce its current stellar mass}. With the $\alpha_\mathrm{sf}$ observationally derived \citep{Da07} at three different redshifts (0.45, 1 and 2) it is possible to derive the SFR at a given $z$ via $SFR = M_\ast/[\alpha_\mathrm{sf} (t_\mathrm{H} - 1 \mathrm{Gyr})]$. In Fig.~\ref{fig:GAMA} the slope is shown for three $z$ values of 0.45, 1 and 2.

\subsection{Observational evidence for a varying galaxy-wide IMF in early-type galaxies}
\label{se:obs2}

While the above studies have been focussed on star-forming, late-type galaxies, recently evidence has also emerged for the galaxy-wide variation of the IMF in early-type or elliptical (E) galaxies. \citet{CMA12} studied a volume-limited sample of 260 E galaxies via integral field spectroscopy and photometry. Using different assumptions on the dark matter halos these galaxies could reside in, from no dark matter to different halo types, \citet{CMA12} found that the SDSS $r$-band mass-to-light ratios of these galaxies do not agree with the assumption of a single slope Salpeter IMF nor with the canonical IMF (see Appendix~\ref{app:IMF}). They conclude that either an extremely bottom-heavy or a very top-heavy IMF in the early universe is necessary to explain the mass-to-light ratios. \citet{VC12} and \citet{CV12}, too, found apparent evidence for a bottom-heavy IMF in ellipticals from NaI, CaII and FeH spectral-line observations. \citet{FPV03} finds a similar trend for bottom-heavy IMFs in the bulges of late-type galaxies. \citet{SMT02} and \citet{CGV03} however disfavour the bottom-heavy IMF as an explanation using Ca observations in E galaxies, naming e.g. the use of the solar Ca abundance ratio not scaled to the metallicity of the E galaxies as a possible solution. The bulges of the Milky Way and M31 have also been reported to have had a top-heavy IMF \citep{BKM07}.

\section{Explaining the IMF variations of galaxies}
\label{se:model}

With all the above compelling evidence for IMF variations in galaxies, is it possible to arrive at a unifying theory which allows quantitative understanding of the observations with a few basic principles?

\subsection{The IGIMF theory}
\label{sub:axioms}
Observationally, IMF variations within the Milky Way have long been considered unlikely or at least rather small \citep{Kr02}. While already noted by \citet{Sc86} the IMF slope of the Milky Way field above 1 $M_\odot$, derived from the present-day stellar mass function, is much steeper ($\alpha_3 \approx$ 2.7) than the Salpeter/Massey slope generally obtained from studies of young star clusters and OB associations in the Milky Way and the Magellanic clouds. The Salpeter/Massey value of $\alpha$ = 2.3 (or 2.35) was and still is widely used in extragalactic astronomy. It was in 2003 that it was shown for the first time that the galaxy-wide IMF should be steeper than the IMFs in individual star-forming regions \citep{KW03}. In 2005 this work was extended to galaxies with different SFRs demonstrating that the galaxy-wide IMF ought to steepen with decreasing SFR \citep{WK05a}. The key assumption of this theory is that stars form in the densest regions of molecular clouds and that the star-formation activity of a whole galaxy is the sum over all these star-forming clumps. That is, stars form in groups or spatially ($\approx$ 1 pc) and temporally ($\approx$ 1 Myr) correlated star formation events (CSFEs), generally termed embedded clusters \citep{LL03,AMB10}. These clusters need not be physically bound and in the highly obscured, deeply embedded phase, it is usually not possible to determine whether they are bound or not. The typical time scale, $\delta t$, to build a system of clusters within a galaxy which statistically samples the mass function of embedded clusters would be about 10 Myr \citep{WKL04}, which is in accordance with the timescale between the formation of molecular clouds and the emergence of new star clusters observed in disk galaxies \citep{ESN04,EKS09} and with CO observations which show that clusters older than 10 Myr do not have large molecular clouds associated with them \citep{LBT89}. In recent years, the following six empirical relations governing galaxy-wide star formation (or ''laws of star-formation'') emerged. These are our axioms upon which the IGIMF theory is based\footnote{We take the axiomatic approach of discovering a theory here: a certain set of observations yield axioms. These can be used as the basis from which a theory that unifies various phenomena and which allows predictions to be made can be formulated.}:

\begin{itemize}
\item[1.] The IMF, $\xi(m)$, within embedded star clusters is canonical (see Appendix~\ref{app:IMF}) for cloud core densities, $\rho_\mathrm{cl}, \lesssim 9.5 \times 10^4 M_\odot$ / pc$^3$, where $\rho_\mathrm{cl}$ = 3$M_\mathrm{cl}$/4$\pi r_\mathrm{h}^3$ and $M_\mathrm{cl}$ is the original molecular cloud core mass in gas and stars, which is for a star-formation efficiency, $\epsilon$, of 33\% \citep{LL03} three times the mass of the embedded cluster, $M_\mathrm{ecl}$, and $r_\mathrm{h}$ is its half-mass radius,
\item[2.] the CSFEs populate an embedded-cluster mass function (ECMF), which is assumed to be a power-law of the form, $\xi_\mathrm{ecl}(M_\mathrm{ecl})$ = $dN$ / $dM_\mathrm{ecl} \propto M_\mathrm{ecl}^{-\beta}$,
\item[3.] the half-mass radii of CSFEs follow $r_\mathrm{h}$ (pc) = 0.1$\times$($M_\mathrm{ecl}$/$M_\odot$)$^{0.13}$ \citep{MK12} yielding $\log_{10}(\rho_\mathrm{cl})$ = 0.61 $\times \log_{10}(M_\mathrm{ecl}/M_\odot)$ + 2.85, in units of $M_\odot$ / pc$^3$,
\item[4.] the most-massive star in a cluster, $m_\mathrm{max}$, is a function of the stellar mass of the embedded cluster, $M_\mathrm{ecl}$, \citep{WK04,WK05b,WKB09,WKP13}, $m_\mathrm{max}$ = $m_\mathrm{max}(M_\mathrm{ecl})$ (e.g. eq.~10 in \citealt{PWK07}),
\item[5.] there exists a relation between the star-formation rate (SFR) of a galaxy and the most-massive young ($<$ 10 Myr) star cluster, $\log_{10}(M_\mathrm{ecl, max}/M_\odot)$ = $0.746 \times \log_{10}(SFR)$ + 4.93 \citep{WKL04}, where the SFR is in units of $M_\odot$ yr$^{-1}$,
\item[6.] the dependence of the IMF slope, $\alpha_3$, of stars above 1 $M_\odot$ on the initial density of the CSFE and metallicity of the CSFE as is given by eq.~\ref{eq:topheavy} below \citep{MKD10}.
\end{itemize}
These axioms make it possible to calculate the galaxy-wide, or integrated galactic stellar initial mass function (IGIMF) \citep{WK05a}, explicitly in dependence of the galaxy-wide SFR and the metallicity,

\begin{eqnarray}
\label{eq:igimf}
\xi_{\rm IGIMF}(m, t) &=& \int_{M_{\rm ecl,min}}^{M_{\rm ecl,max}(SFR(t))} \xi(m \le m_{\rm max}(M_{\rm ecl})) \cdot \nonumber \\
&&\xi_{\rm ecl}(M_{\rm ecl})\,dM_{\rm ecl}.
\end{eqnarray}
Here $\xi(m\le m_{\rm max})~\xi_{\rm ecl}(M_{\rm ecl})~dM_{\rm ecl}$ is the stellar IMF, with $m_\mathrm{max}$ limited by $M_\mathrm{ecl}$ (above axiom 4), contributed by $\xi_{\rm ecl}~dM_{\rm ecl}$ clusters with stellar mass in the interval $M_\mathrm{ecl}$, $M_\mathrm{ecl} + dM_\mathrm{ecl}$. While $M_{\rm ecl,max}$ follows from the SFR-$M_\mathrm{ecl,max}$-relation (above axiom 5), $M_{\rm ecl,min}\,=\,5\,M_{\odot}$ is generally adopted, corresponding to the smallest known CSFE, namely the individual groups of very young ($\lesssim$ 1 Myr) stars in Taurus-Auriga \citep{KB03,KM10}. Due to the dependence of $M_\mathrm{ecl, max}$ on the SFR the IGIMF depends on the SFR of a galaxy. 

In the case of starbursts, it has been recently found by \citet{DKB09,DKP12} and \citet{MKD10} from an analysis of globular clusters and ultra-compact dwarf-galaxies that the IMF within CSFEs becomes top-heavy under very large star-formation-rate densities and that it can be described by,

\begin{equation}
\label{eq:topheavy}
\alpha_3  = \left\{\begin{array}{ll}
\hspace*{0.25cm} +2.3&, x < -0.87\\
-0.41 \times x + 1.94&, x \ge -0.87,\\
\end{array}\right.
\end{equation}
\noindent
with $x$ = $-$0.14 [Fe/H] + 0.99 $\log_{10} (\rho_\mathrm{cl}/(10^6 M_\odot \mathrm{pc}^{-3}))$. We here correct a minor error in the IMF formula of \citet{MKD10} where erroneously $x \ge 0.87$ in their eq.~14 but it should instead read $x \ge -0.87$. \citet{MKD10} find a limit of $9.5 \cdot 10^4 M_\odot$ / pc$^3$, above which the IMF in CSFEs becomes top-heavy. This translates into a cluster mass of $M_\mathrm{ecl} >$ 2.7 $\cdot\,10^5\,M_\odot$ when using the radius-$M_\mathrm{ecl}$ relation for embedded clusters from \citet{MK12} (axiom 3 above) and assuming a star-formation efficiency of 33\%. Solar metallicity is assumed for all calculations in this work.

A more comprehensive overview of  the theoretical and observational background of the IGIMF is given in \citet{KWP13}.

With the IGIMF theory it was not only possible to predict the H$\alpha$-flux to UV-flux dependence of the SFR of a galaxy \citep{PWK07,PWK09}, as has been observed by \citet{MWK09} and \citet{LGT09}, but the IGIMF reproduces these flux variations readily without any parameter adjustments. Furthermore, the IGIMF very naturally reproduces the observed mass-metallicity-relation of galaxies \citep{KWK05} as well as the differences in metallicity  between disk and bulge stars in the Milky Way \citep{CRM10} and reduces the need for galaxy down-sizing \citep{RCK09}.

\subsection{Star forming galaxies}
\label{sub:late}

\begin{figure}
\begin{center}
\includegraphics[width=9cm]{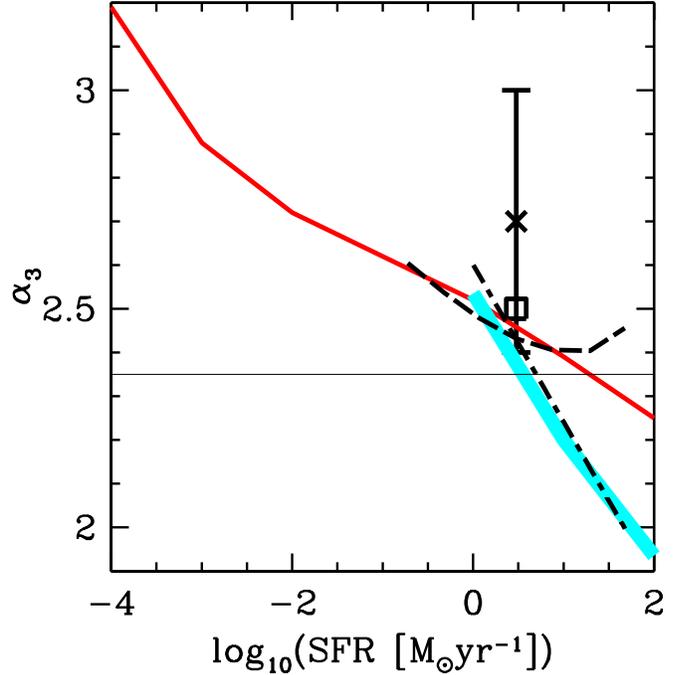}
\vspace*{-1.5cm}
\caption{Like Fig.~\ref{fig:GAMA} but additionally including as the thick red line the IGIMF slope which readily describes the \citet{LGT09} results for dwarf galaxies assuming $\beta$ = 2 = constant. This IGIMF model is extended into the starburst regime as described in \citet{KWP13} and based on axioms 1--6. Assuming additionally that $\beta$ varies according to eq.~\ref{eq:beta} yields the IGIMF behaviour shown as the thick grey line.}
\label{fig:GAMA2}
\end{center}
\end{figure}

Fig.~\ref{fig:GAMA2} shows the IGIMF model with the parameters used to explain the H$\alpha$-flux to UV-flux  variation and the top-heavy extension as described above (axioms 1--6) as a thick solid line together with the constrains from \citet{HG07} and \citet{GHS10}.

The recent observational analysis by \citet{GHS10} indicate a stronger flattening of the galaxy-wide mass function slope with increasing SFR at large SFRs than calculated for the IGIMF as based on axioms 1--6. In \citet{WK07a} the possibility of a systematic variation with SFR of the lower mass limit of the ECMF, $M_\mathrm{ecl, min}$, and/or of the slope of the ECMF, $\beta$, were considered and these possibilities are reconsidered here as potential solutions for the observed stronger flattening. Generally, a variation of the ECMF such that it becomes increasingly top-heavy with increasing SFR would be in accord with the general expectation (e.g.~the Jeans mass increasing with increasing ambient temperature).Very little is known about $M_\mathrm{ecl, min}$ as it is not straightforward to define what might be the smallest possible CSFE \citep{LL03}. Considerable scatter is found for $\beta$ \citep[1.8 - 2.5,][]{Larsen2009} and so far no clear indication for a variation of $\beta$ with the SFR has been discovered. But the ECMF of the Antennae interacting galaxies, which have a high SFR $\approx$ 20 $M_\odot$ yr$^{-1}$ \citep{ZFW01}, seems to be flatter than the cluster mass functions in normal spirals \citep[fig.~6 in][]{Larsen2009}. Direct measurement might be difficult as clusters are known to be destroyed rapidly \citep{KB02,BK03a} and it is not clear whether this process is mass dependent or not.

The \citet{GHS10} study uses initially a single slope Salpeter IMF to determine the SFRs before introducing a variation of the slope. This prompts the question if their results can be directly used with the IGIMF which is based on the canonical IMF as described in Appendix~\ref{app:IMF}. However, the similarity of the \citet{GHS10} results with studies using different methods and IMFs, like \citet{Da07} and \citet{BKM07}, gives confidence that the impact of the differences cannot be very large. Furthermore, \citet{FVR13} studied the impact of IMF variations on recovered galaxy properties, like the total mass, for single-slope and two-slope IMFs and found that only for very steep IMFs the recovered galaxy mass varies strongly while for flat slopes the variation is very limited.

The small $\alpha_3$ values at high SFRs \citep[$\alpha_3$ $\approx$ 2 at SFR $\approx$ 10$^2$ $M_\odot$ yr$^{-1}$ from][]{Da07,GHS10} shows a limitation in the present IGIMF theory as based on axioms 1 - 6 in that $\alpha_3$ cannot reach such small values at this SFR. To address this we need to introduce an additional axiom 7 to the IGIMF, which can be achieved by either adjusting the power-law index, $\beta$, of the ECMF or the minimum cluster mass with SFR.

We therefore suggest the following relation between $\beta$ and the SFR for 1 $\le$ SFR/($M_\odot$ yr$^{-1}$) $\le$ 50, in order to reproduce the \citet{GHS10} constrains shown in Fig.~\ref{fig:GAMA2},

\begin{equation}
\label{eq:beta}
\beta = \left\{\begin{array}{ll}
\hspace*{0.25cm} 2.00&, SFR < 1 M_\odot~\mathrm{yr}^{-1}\\
-0.106 \times \log_{10}(SFR) + 2.00&, SFR \ge 1 M_\odot~\mathrm{yr}^{-1},\\
\end{array}\right.
\end{equation}
with $\beta$ being the slope of the ECMF and SFR is in units of $M_\odot$ yr$^{-1}$. The equation is arrived at by varying $\beta$ for a given SFR and calculating the resulting IGIMF slope. When this slope is within 0.01 dex of the SFR-$\alpha_3$-relation of \citet{GHS10} the value is used for the fit. While the \citet{GHS10} data cover SFRs only up to 50 $M_\odot$ yr$^{-1}$, we apply eq.~\ref{eq:beta} to large SFRs.

An alternative possibility would be $\beta$ = 2 = constant but a changing lower mass limit of the ECMF, $M_\mathrm{ecl, min}$, and the following fit results in the same IGIMF changes as eq.~\ref{eq:beta},

\begin{displaymath}
\log_{10}(M_\mathrm{ecl,min}/M_\odot) = 
\end{displaymath}
\begin{equation}
\label{eq:meclmin}
 \left\{\begin{array}{ll}
0.70 &, SFR < 1 M_\odot~\mathrm{yr}^{-1},\\
1.94 * \log10(SFR) + 0.70&, SFR \ge 1 M_\odot~\mathrm{yr}^{-1}.\\
\end{array}\right.
\end{equation}
Both descriptions have been tested and deliver identical results but, as stated above, variations of $M_\mathrm{ecl, min}$ are at the moment virtually non-constrained by observations. It may become possible to obtain constrains with in-depth HST observations or with the upcoming JWST.

\begin{figure}
\begin{center}
\includegraphics[width=9cm]{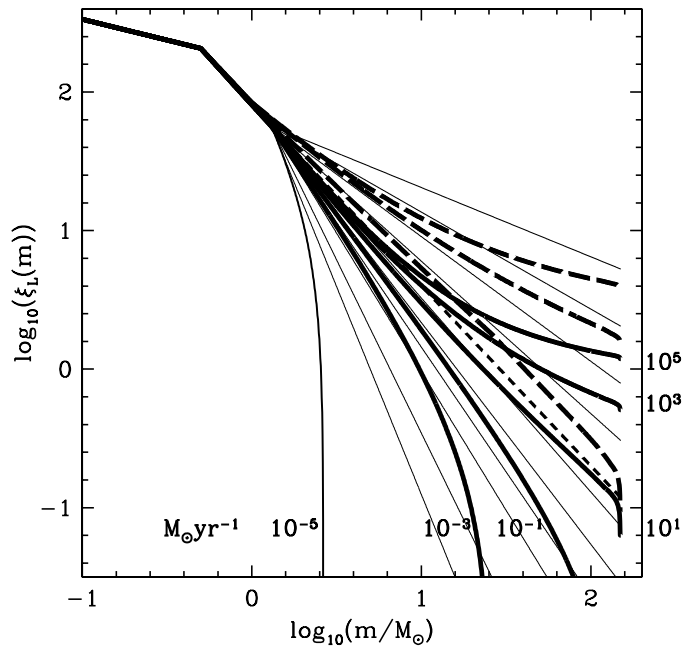}
\vspace*{-2.5cm}
\caption{The dependence of the logarithmic IGIMF (eq.~\ref{eq:igimf}) on the SFR of a galaxy. The IGIMF is normalised to the same values at $m$ $<$ 1 $M_\odot$. The IGIMF is plotted with thick solid lines. It uses the canonical IMF which becomes top-heavy at solar metallicity embedded-star-cluster densities (gas + stars) $\rho_\mathrm{cl} > 9.5 \times 10^4~M_\odot / \mathrm{pc}^3$ (eq.~\ref{eq:topheavy}), with a constant ECMF, $\beta$ = 2, $M_\mathrm{ecl,min}$ = 5 $M_\odot$ and the $m_\mathrm{max}-M_\mathrm{ecl}$ relation (axiom 4). The corresponding SFRs for the thick solid lines are indicated as numbers in the plot (SFR = $10^{-5}, 10^{-3}, 10^{-1}, 10^1, 10^3, 10^5$) and are in units of $M_\odot$ yr$^{-1}$ (from left to right). The case when $\beta$ is dependent on the SFR for SFR $\ge$ 1 $M_\odot$ yr$^{-1}$ (eq.~\ref{eq:beta}) is shown with thick long-dashed lines for SFRs of $10^1, 10^3, 10^5$ $M_\odot$ yr$^{-1}$ from bottom to top. The thin lines are IMFs with different power-law indices, $\alpha'$, for m $>$ 1.3 $M_\odot$. $\alpha'$ = 1.5, 1.7, 1.9, 2.1, 2.3, 2.4, 2.6, 2.8, 3.0, 3.5, 4.0 (top to bottom), whereby the canonical value $\alpha'$ = 2.3 = $\alpha_3$ is shown as the thick  short-dashed line. Adopted from fig.~35 from \citet{KWP13}. Note that the IGIMF has $\alpha_3^{'}$ $\approx$ 2.6 for SFR = 1 $M_\odot$ yr$^{-1}$ in agreement with the \citet{Sc86} determination of the Galactic-field IMF.}
\label{fig:igimf}
\end{center}
\end{figure}

While it is possible to explain the \citet{GHS10} results with a varying $\beta$ or $M_\mathrm{ecl, min}$ (our 7th axiom), other explanations have been proposed. For example a diffuse mode of star-formation with a truncated or a variable IMF \citep{MHL95,L04} might also be able to reproduce the IMF slope behaviour.

The resulting form of the IGIMF calculated from the above seven axioms is shown in Fig.~\ref{fig:igimf} in dependence of the galaxy-wide SFR.

\subsection{Elliptical galaxies}

With the IGIMF theory as defined by eq.~\ref{eq:igimf} and axioms 1--7 it is possible to account for the observationally determined $\alpha_3$ variations for late-type galaxies that have $10^{-5} \le$ SFR / ($M_\odot$ yr$^{-1}$) $\le 10^4$ (Fig.~\ref{fig:GAMA2}). Can the same IGIMF theory also account for the properties of E galaxies?\\

\begin{figure}
\includegraphics[width=8cm]{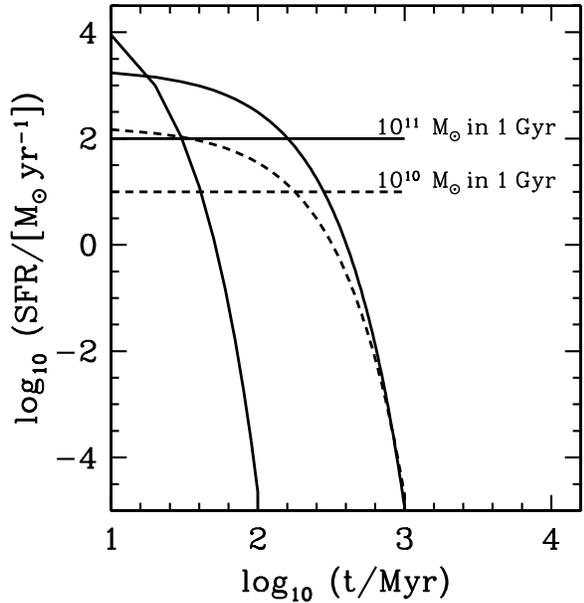}
\vspace*{-1.5cm}
   \caption{Typical star-formation histories used for the calculation of the IGIMF models for E galaxies. The solid lines refer to galaxies which form in total 10$^{11}$ $M_\odot$, either constantly over 1 Gyr or exponentially declining over 100 Myr or 1 Gyr. The dashed lines are SFHs for a total mass of 10$^{10}$ $M_\odot$ formed either with a constant SFR over one Gyr or with an exponentially declining SFR over 100 Myr . The masses shown are the total masses prior to stellar evolution reducing the present-day masses of the galaxies which has been taken into account in the calculations of the M/L-ratios.}
              \label{fig:sfh}
\end{figure}

\begin{figure*}
\includegraphics[width=8.5cm]{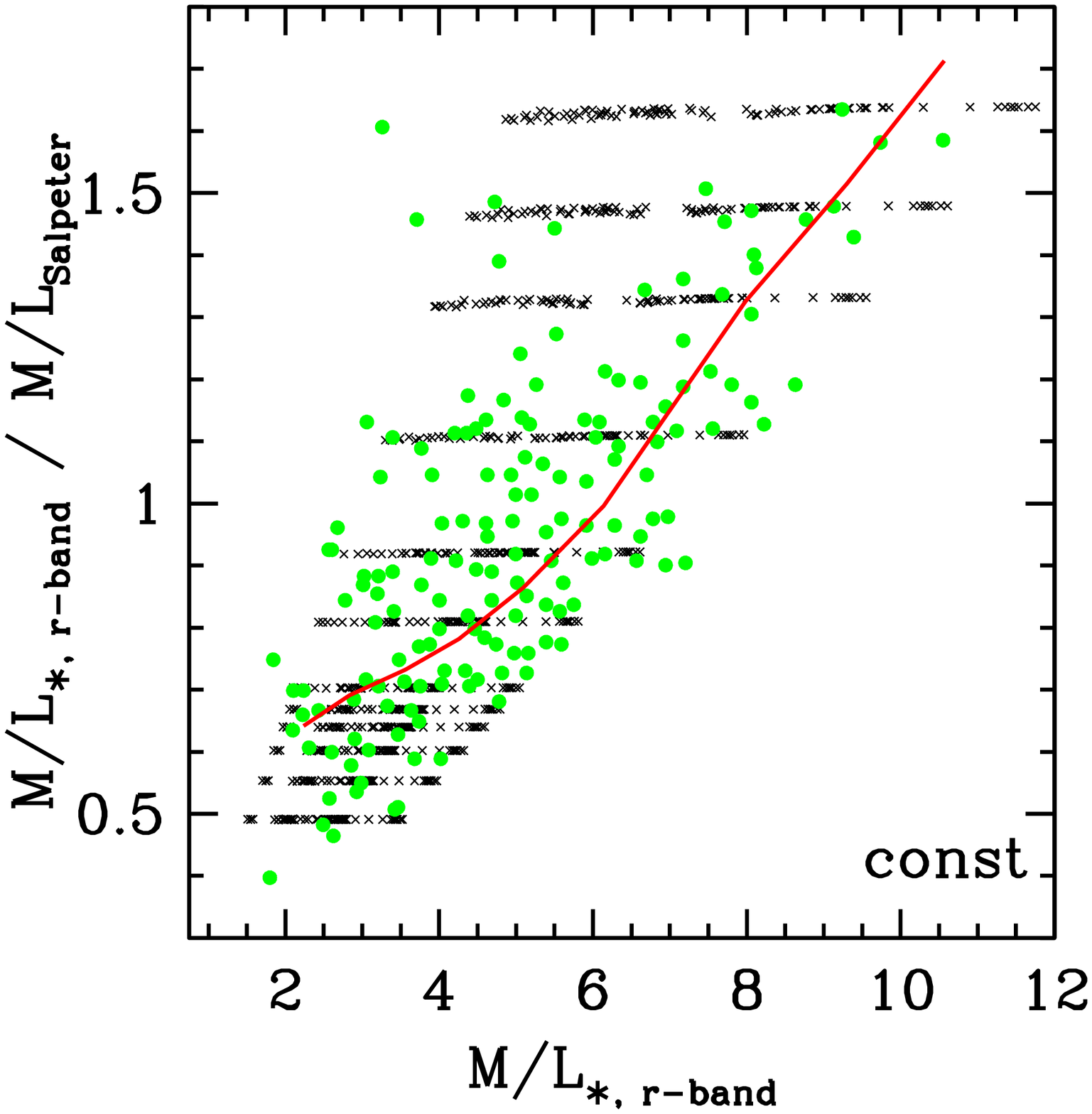}
\includegraphics[width=8.5cm]{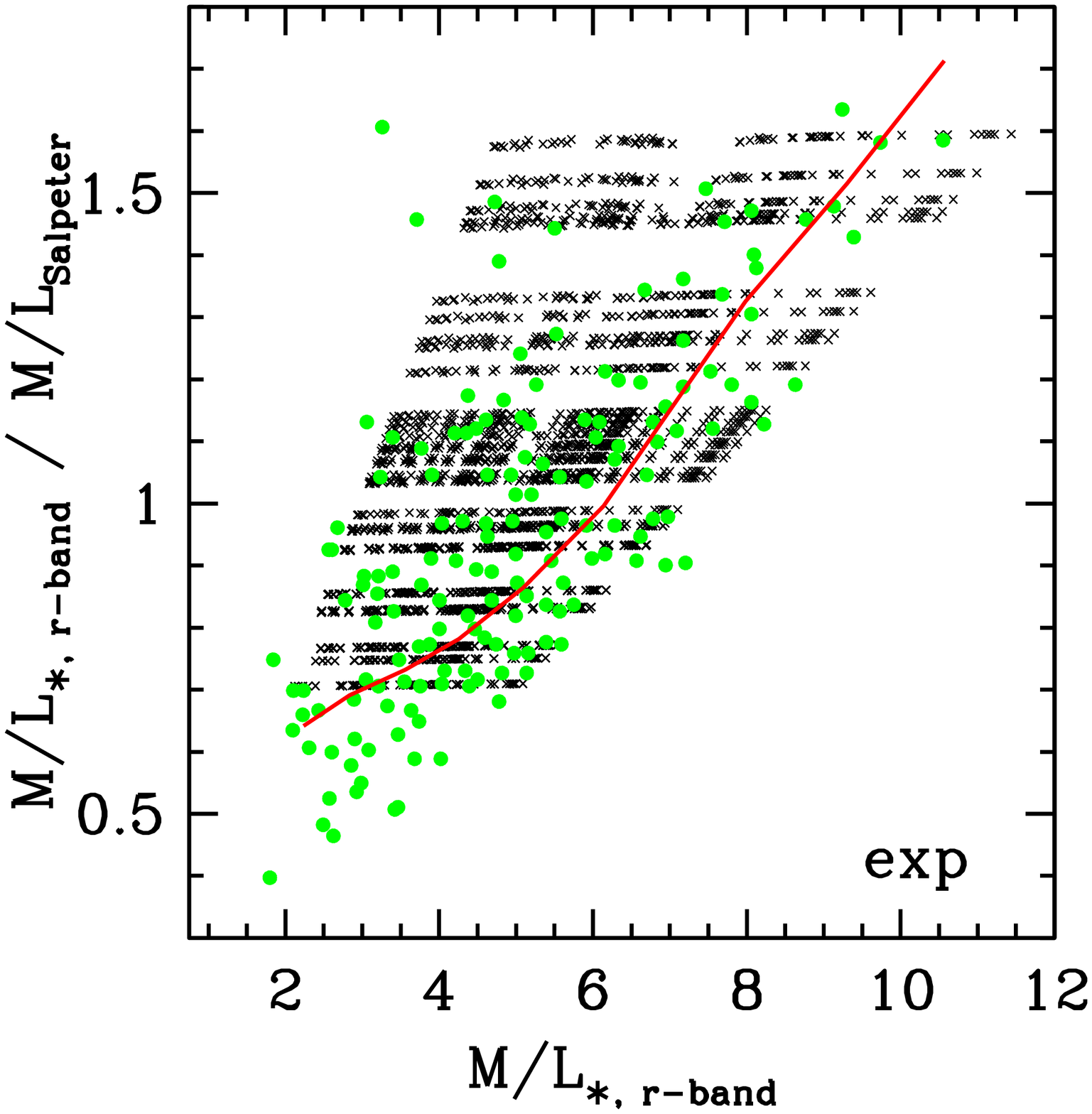}
\vspace*{-1.5cm}
   \caption{The $r$-band mass-to-light ratios divided by the mass-to-light ratios when assuming a constant Salpter IMF against $r$-band mass-to-light ratios. The crosses in both panels are calculated using the IGIMF theory, while the solid dots in both panels are the observational results of a sample of 260 galaxies by \citet{CMA12}. The solid line is a {\it loess} smoothed version of the observational data. In the panel named 'const' the observational data are overlaid with IGIMF models (crosses) with different constant SFRs ($10^{-5}$ to $10^{3.5}$ $M_\odot$ yr$^{-1}$, from bottom to top) for 1 Gyr. In the panel 'exp' are shown IGIMF models (crosses) with exponentially declining SFRs for 100 Myr and 1 Gyr with $M_\mathrm{tot} = 10^8$ to $10^{12} M_\odot$ (from bottom to top). Examples for the SFHs are shown in Fig.~\ref{fig:sfh}. The lowest M/L ratios, which are not covered by the exponential models, are the lowest mass objects. These might have more extended star-formation histories which are not covered by our models. We extracted the M/L$_{\ast, \mathrm{r-band}}$ / M/L$_{\mathrm{Salpeter}}$ and the M/L$_{\ast, \mathrm{r-band}}$ values from Figure 2 of \citet{CMA12} using the tool PlotDigitizer.}
              \label{fig:m2l}
\end{figure*}

\begin{figure}
\includegraphics[width=8cm]{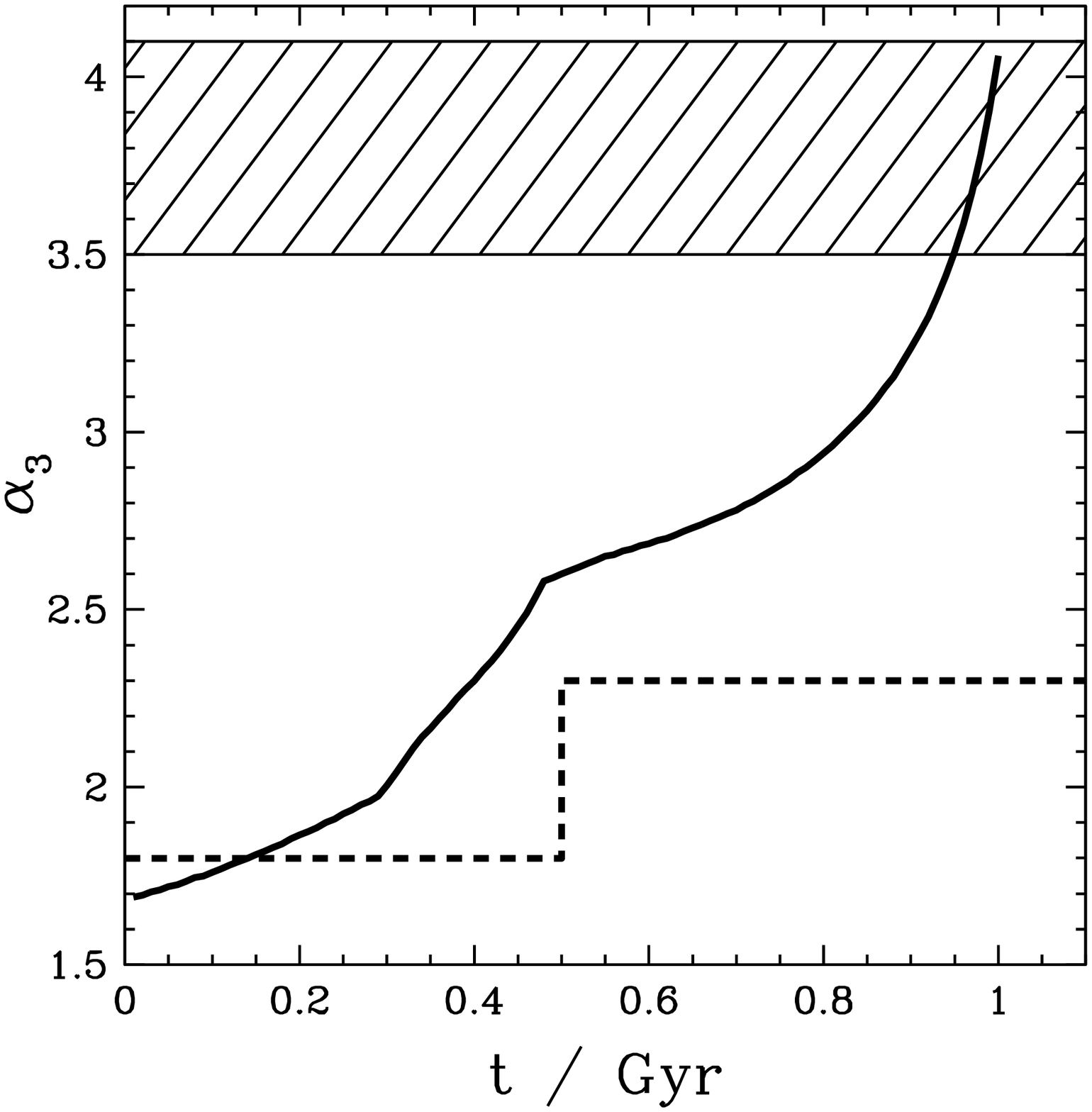}
\vspace*{-1.5cm}
   \caption{Variation of the IGIMF index $\alpha_3$ with time during the formation of elliptical galaxies. The solid line shows $\alpha_3$ for IGIMF model III for a galaxy with 10$^{12}$ $M_\odot$ and an exponentially declining SFH over 1 Gyr as show in Fig.~\ref{fig:sfh}, while the dashed line is the \citet{VCP96} model, which uses a short burst (0.2 to 1 Gyr) and then an exponentially declining star-formation rate. The shaded area marks the range of IMF power-law indices found by \citet{FBR13} for ellipticals with central velocity dispersions between 250 and 300 km/s derived from fitting line-indices for different tracers with single stellar population models and the {\sc Starlight} code \citep{CMS05}.}
              \label{fig:slope}
\end{figure}

Elliptical galaxies are known to be very old and to typically have formed as major bursts with 10 $\lesssim$ SFR / ($M_\odot$ yr$^{-1}$) $\lesssim$ 10$^4$. According to the IGIMF theory they ought to have produced a large fraction of stellar remnants (Fig.~\ref{fig:igimf}).
To compare the \citet{CMA12} results for E galaxies with IGIMF models it is necessary to calculate the M/L-ratios for the models. As available stellar population synthesis models do not allow for a variable IMF these models had to be calculated using our own code. In order to do so, the mass axis of the IGIMF is divided into 2000 logarithmic mass bins. As a lower mass limit 0.1 $M_\odot$ is used and 100 $M_\odot$ is the upper limit. The centre of each bin is treated as a single star and evolved over 15 Gyr in 1 Myr time-steps using stellar evolution models (references for the models are listed in Tab.~\ref{tab:models} and the initial-final mass relation used is detailed in Tab.~\ref{tab:rem}). At each step the effective temperature, $T_\mathrm{eff}$, and surface gravity, $\log_{10} g$, of the models is used to locate the appropriate stellar atmosphere model \citep{HAB99}. The resulting spectrum is then integrated over the SDSS $r$-band filter curve \citep{GCR98} to obtain the fraction of the total luminosity of the stars in the SDSS $r$-band. The resulting $L_r$ is averaged over a 10 Myr time interval to match the $\delta t$ = 10 Myr time-scale necessary to fully populate the ECMF in the IGIMF models. With the IGIMF models the number of stars formed per $\delta t$ in each mass bin is calculated and multiplied with $L_r$. The luminosity and mass of the stars are updated every time step according to the age of each population. Two different types of star-formation histories (SFH) are used (examples are shown in Fig.~\ref{fig:sfh}). The SFR is either assumed to be constant between $10^{-5}$ $M_\odot$ yr$^{-1}$ and $10^{4}$ $M_\odot$ yr$^{-1}$ over 1 Gyr (left panel of Fig.~\ref{fig:m2l}) or exponentially declining on 100 and 1000 Myr exponential time scales (right panel of Fig.~\ref{fig:m2l}). {\it This leads to an IGIMF which is top-heavy early-on (low metallicity, high SFR) and which evolves towards an increasingly top-light and bottom-heavy IGIMF as the metallicity increases and SFR decreases (Fig.~\ref{fig:slope})}. It should be noted here that \citet{VCP96} and \citet{VPB97} already proposed an IMF change in E galaxies in a two-phase model. They start with a flat IMF early on ($<$ 0.5 Gyr) which later turns into a Salpeter-like one, in order to achieve significantly improved fits for various line-strengths and colours for ellipticals. The \citet{VCP96} model, our IGIMF model III (see \S~\ref{se:res}) and the observational results by \citet{FBR13} are compared in Fig.~\ref{fig:slope}. It is also important to notice that a purely bottom-heavy IMF for giant elliptical galaxies as proposed by \citet{VC10} is at odds with the chemical evolution of these objects as they have solar or super-solar metallicities which are impossible to reproduce with very steep IMFs. However, a top-heavy galaxy-wide IMF burst with a later-on bottom-heavy galaxy-wide IMF works well \citep{WFV13}.

\begin{table*}
\caption{\label{tab:models} Stellar evolution models used in this work.}
\begin{tabular}{ccc}
evolutionary&mass range&reference\\
phase&$M_\odot$&\\
\hline
MS&0.08 - 0.15&\citet{BHS93,BMH97}\\
MS&0.16 - 0.8&\citet{HPT00}\\
MS&0.9 - 10.0&\citet{CPC07}\\
RGB, HB, E-AGB&0.9 - 10.0&\citet{CPC07}\\
TP-AGB&0.5 - 10.0&\citet{PCS06}\\
all phases&10.0 - 120.0&\citet{MM03}\\
White Dwarf remnants&$<$ 7.0&\citet{FWL05}\\
Neutronstar \& Black Hole remnants&7.0 - 120.0&\citet{WHW02}\\
\end{tabular}

PMS: pre-main-sequence; MS: main-sequence; RGB: red giant branch; HB: horizontal branch; 
E-AGB: early asymptotic giant branch; TP-AGB: thermal-pulse asymptotic giant branch;
\end{table*}

\begin{table}
{\centering
\caption{\label{tab:rem} Remnant types and final masses, $m_{\rm fin}$,
  for massive stars, given their inital mass, $m_\mathrm{ini}$, in $M_\odot$. The
  fitting formulae have been developed for the models shown in fig.~16
  in \citet{WHW02}. C/O denotes a carbon and oxygen core white dwarf
  (wd), O/Ne an oxygen and neon core wd, NS is a neutron star and BH a
  black hole.}
\begin{tabular}{ccl}
\hline
$m_\mathrm{ini}$&remnant&$m_\mathrm{fin}$\\
{[$M_{\odot}$]}&type&[$M_{\odot}$]\\
\hline
0.5 to 7.0&C/O wd&$m_{\rm fin} = 0.1004 m_\mathrm{ini} + 0.4344$$^{a}$\\
7.0 to 8.7& O/Ne wd&$m_{\rm fin} = 0.1088 m_\mathrm{ini} + 0.3757$\\
8.7 to 10.0& NS&$m_{\rm fin} = 0.2915 m_\mathrm{ini} - 1.2124$\\
10.0 to 12.5& NS&$m_{\rm fin} = 0.0648 m_\mathrm{ini} + 1.0550$\\
12.5 to 21.8& NS&$m_{\rm fin} = 1.865$\\
21.8 to 24.8& NS/BH$^b$&$m_{\rm fin} = 0.1980 m_\mathrm{ini} - 2.4514$\\
24.8 to 30.0& BH&$m_{\rm fin} = 1.2087 m_\mathrm{ini} - 27.5162$\\
30.0 to 36.0& BH&$m_{\rm fin} = 0.1193 m_\mathrm{ini} + 5.1640$\\
36.0 to 52.8& BH&$m_{\rm fin} = -0.2780 m_\mathrm{ini} + 19.4671$\\
52.8 to 57.8& BH&$m_{\rm fin} = -0.3510 m_\mathrm{ini} + 23.3218$\\
57.8 to 93.9& NS/BH$^{c}$&
$m_{\rm fin} = \frac{81.205}{\sqrt{(2\,\pi)}\,17.645}\,e^{0.5\left(\frac{m_\mathrm{ini}-75.4}{17.645}\right)^{2}}$\\
93.9 to 100.0& BH&$m_{\rm fin} = 0.2205 m_\mathrm{ini} - 17.5221$\\
100.0$^{+}$& BH&$m_{\rm fin} = 0.1323 m_\mathrm{ini} - 8.7035$\\
\hline
\end{tabular}
}

$^{a}$ Formula from \citet{FWL05}.\\
$^{b}$ NS if $m_\mathrm{ini}\,<$ 23.7 $M_{\odot}$.\\
$^{c}$ NS if $m_{\rm fin}\,<$ 2.14 $M_{\odot}$.
\end{table}

In Fig.~\ref{fig:m2l}, the $r$-band IGIMF M/L-ratios, divided by M/L ratios obtained by the same procedure as described above but using a constant Salpeter-IMF from 0.1 to 100 $M_\odot$ are plotted in dependence of the IGIMF M/L ratio. For the IGIMF models eq.~\ref{eq:beta} is applied to determine $\beta$ for the given SFR. As E galaxies are dominated by old populations, in Fig.~\ref{fig:m2l} only the M/L ratios for stellar populations with ages between 10 and 14 Gyr are plotted. The \citet{CMA12} results are shown as light grey dots in Fig.~\ref{fig:m2l}, assuming no dark matter within the effective radii of the galaxies. But because only very little dark matter is expected inside these radii the dependence on the type and shape of a potential dark matter halo is negligible \citep{CMA12}. The red solid line is the mean of the observations. 

\begin{figure}
\includegraphics[width=8.5cm]{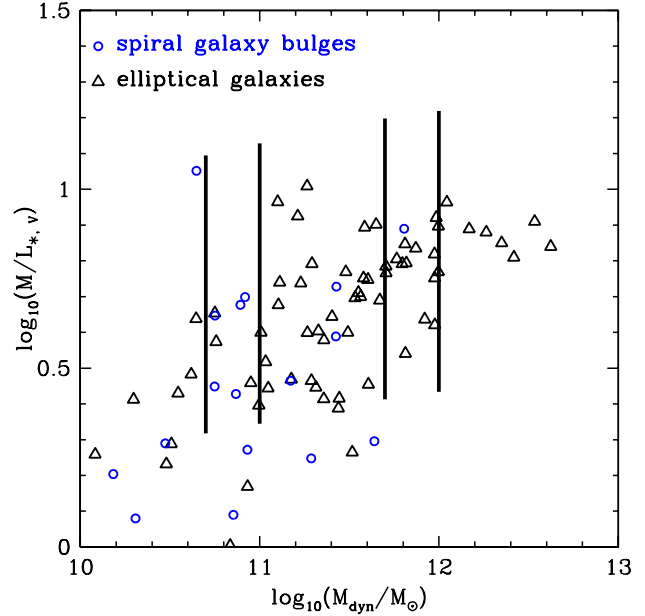}
\vspace*{-2.0cm}
   \caption{The $V$-band mass-to-light ratios in dependence of dynamical masses for observed  elliptical galaxies (triangles) and bulges of spiral galaxies (circles) from a compilation in \citet{DHK08}. Additionally are plotted as solid lines the range of mass-to-light ratios of models with an exponential SFH with an exponential time scale of 1~Gyr evaluated at ages between 6 and 12 Gyr. Generally, the lower end of the M/L ratios correspond to an age of 6 Gyr age and the upper end to 12 Gyr, though the ratios are not exactly linear in time. As the observations tend to have lower mass-to-light ratios than the models at lower galaxy masses it could be that these galaxies had more extended SFHs. A similar trend has been found in chemical evolution models of galaxies able to reproduce the observed [$\alpha$/Fe] vs. velocity dispersion relation \citep{RCK09}.}
              \label{fig:m2lV}
\end{figure}

To compare the model results with a different set of observations, the \citet{DHK08} compilation of $V$-band M/L ratios of elliptical galaxies (triangles) and bulges of spiral galaxies (circles) are plotted versus the dynamical mass of the systems together with the model results (solid lines) in Fig.~\ref{fig:m2lV}. For the lines it was assumed that all stars formed between 6 and 12 Gyr ago within 1 Gyr with an exponentially declining SFR. The agreement between models and observations is reasonable, considering the model and observational uncertainties, and the trend of declining M/L ratio with decreasing mass is reproduced. The faster decline of the observed M/L$_\mathrm{V}$ values with decreasing mass can be seen as possible evidence that star-formation histories were more extended in lower mass systems than in massive E galaxies, as is also reported for example by \citet{RCK09} or \citet{RFP10}.

The IGIMF calculations shown here thus cover the range of the observations very well. It can be concluded that the IGIMF theory based on the seven axioms of \S~\ref{sub:axioms} and \S~\ref{sub:late} explains the \citet{CMA12} results. Thus, by having constrained $\beta$ = fn(SFR) (eq.~\ref{eq:beta}) using late-type star-forming galaxies we arrive at a consistent description of early-type galaxies without needing any further adjustments. Nevertheless, additional tests of the IGIMF-theory are required.

\subsection{The Milky Way (MW) case}
\label{se:MW}

In the MW many star clusters and their IMF are well studied but the galaxy-wide IMF of the MW is not that well constrained. In his seminal work \citet{Sal55} found a power-law index of $\alpha_\mathrm{Salpeter}$ = 2.35 for stars in the field with masses between 0.4 and 10 $M_\odot$. Later, \citet{Sc86} with more sophisticated modelling and better star count data, corrected the field power-law index above a few tenth of a $M_\odot$ to $\alpha_\mathrm{Scalo}$ = 2.7 but with a large uncertainty of 0.3 dex \citep{KTG93}. Using the IGIMF as defined by axioms 1--7 results, for a SFR of 1 $M_\odot$ yr$^{-1}$, in a power-law index for stars above 1 $M_\odot$ of $\alpha_3$ = 2.6, somewhat steeper than the \citet{Sal55} value and well consistent with the \citet{Sc86} determination.

No good determination of the IMF of the stellar population of the MW halo exists but recent work on the bulge of the MW indicates a difference in the IMF in comparison with the MW field.  \citet{BKM07} used chemical evolution modelling to investigate how to reproduce the metallicity distribution of stars in the MW bulge. They found that a model with a rapid star-formation (formation period of 0.1 Gyr) and an IMF power-law index of $\alpha_\mathrm{Balero}$ = 1.95 for stars above 1 $M_\odot$ is necessary to explain the observed metallicities. Assuming that the MW bulge comprises about 10\% of the MW stellar population ($M_\mathrm{bulge}$ = 10$^{10} M_\odot$) and using the 0.1 Gyr formation period, the IGIMF theory yields a power-law index $\alpha_3$ = 1.94, in excellent agreement with the \citet{BKM07} value.

Thus, the galaxy-wide IMF of the MW is very well described by the IGIMF theory.
\section{The expected number of OB stars from the IGIMF as a function of the SFR}
\label{se:res}

As a test of the IGIMF theory it is possible to predict the number of O and B stars for whole galaxies for a given SFR. These numbers then need to be compared to observational results of galaxies with a range of SFRs.

Here three different IGIMF models are compared with a canonical IMF model:\\

$
\begin{array}{ll} 
\mathrm{IGIMF~model~I:}&\mathrm{axiom~1} - \mathrm{5~with}~\beta = 2,\\
\mathrm{IGIMF~model~II:}&\mathrm{axiom~1} - \mathrm{6~with}~\beta = 2,\\
\mathrm{IGIMF~model~III:}&\mathrm{axiom~1} - \mathrm{7~with}~\beta = fn(\mathrm{SFR}).\\
\nonumber
\end{array}
$

\noindent
Note that the parameters of the IGIMF model are not freely adjusted but are based on the axioms of \S~\ref{sub:axioms} and \S~\ref{sub:late} which follow from observational constrains \citep[see also][]{KWP13}. 

The number of O stars for each model is relatively easy to calculate for a given SFR as the maximum lifetime of O stars is shorter than $\delta t$ = 10 Myr, which is the typical timescale for a population of star clusters to hatch from their embedded phase \citep{WKL04,ESN04,EKS09}. Therefore, with $m_1$ = 18 $M_\odot$ and $m_2$ = 150 $M_\odot$ as limits for O stars, the number of O stars for the different SFRs are listed in Tab.~\ref{tab:numbers}, being calculated as

\begin{equation}
\label{eq:N}
N_{\ast} = \int_{m_1}^{m_2} \xi(m) dm.
\end{equation}
When a new period of star-formation starts after $\delta t$, all the O stars formed in the previous period are gone as supernovae. 

To calculate the number of B stars is more complicated as B stars have life times up to several hundred Myr. In order to derive this number a steady state of forming and dying B stars is assumed. This is done by dividing the range of masses of B stars ($m_1$ = 3 to $m_2$ = 18 $M_\odot$) into 15 bins (each 1 $M_\odot$ wide) and taking the mean time, $\tau_\mathrm{B}$, the stars in each bin are of spectral type B from stellar evolution models \citep{CPC07,MM03}. The $\tau_\mathrm{B}$ are then multiplied by the number of stars in each mass bin derived from using eq.~\ref{eq:N} and the IGIMF models to obtain the total number of B stars. For example, if $\tau_\mathrm{B}$ is 200 Myr for a given mass bin, and the number of B stars in this bin formed during $\delta t$ = 10 Myr is 1000, the total equilibrium contribution to the number of B stars from this bin would be 1000 $\times$ 20 = 20000. The total number of B stars is arrived at by summing the contributions of each mass bin.

These predicted relative numbers of B and O type stars for a given SFR are shown in Table~\ref{tab:numbers}.
In the first column are listed the different SFRs. In the second column are the absolute numbers of B stars to exist in a galaxy with the given SFR assuming an invariant canonical IMF. These values are the mean numbers to be expected from the canonical IMF. Between individual galaxies a certain level of statistical variation is to be expected. The same is given in the third column, though for O stars instead of B stars. Columns 4 and 5 give the relative B and O star numbers, respectively, for the IGIMF with a constant $\alpha_3$ of 2.35 for the IMF within the star clusters and a constant $\beta$ of 2 for the ECMF (IGIMF model I). In columns 6 and 7 are given the relative numbers of B and O stars formed when assuming that the IMF of very massive clusters is top-heavy (see eq.~\ref{eq:topheavy}) leading to a top-heavy IGIMF (IGIMF model II) but the ECMF slope still has a constant value of 2. For the columns 8 and 9 (IGIMF model III) additionally $\beta$ varies with SFR according to eq.~\ref{eq:beta}. The values in column 4 to 9 are all relative to the canonical IMF numbers. In order to get the actual numbers, column 4, 6 and 8 have to be multiplied by column 2, while columns 5, 7 and 9 need to be multiplied by column 3. For example, for a galaxy with a SFR of 10$^{-4}$ $M_\odot$ yr$^{-1}$, about 385 B stars and 3 O stars are to be expected if the IGIMF in that galaxy is identical to the canonical IMF. If, instead, the IGIMF model I with $\beta$ = 2 is applied, there should be no O stars and 212 B stars (385 $\times$ 0.55). Those cases where the results of the top-heavy IGIMF models differ from the non-top-heavy IGIMF are marked in boldface. 

\begin{table*}
\caption{\label{tab:numbers} The expected number of O and B stars in a galaxy in dependence of the SFR of the galaxy. Columns 2 and 3 are the expected number of O and B stars for the canonical invariant IMF between the mass limits of 0.1 and 100 $M_\odot$. Columns 4 to 9 are the numbers for 3 different IGIMF models. Columns 4 and 5 are for IGIMF model I with constant $\alpha_3$ = 2.35 for the IMF and constant $\beta$ = 2 for the ECMF. In columns 6 and 7 are the IGIMF model II that are top-heavy because $\alpha_3$ = fn($\rho_\mathrm{ecl}$) as described by eq.~\ref{eq:topheavy}. And for columns 8 and 9 (model III), $\alpha_3$ varies as for columns 6 and 7 but also $\beta$ = fn(SFR). In columns 4, 6 and 8 the B star numbers are given in fractions of column 2, while in columns 5, 7 and 9 the O star numbers are given in fractions of column 3. In order to, for example, calculate the actual number of expected O stars for IGIMF I with constant $\alpha_3$ and $\beta$ for  a SFR of 1 $M_\odot$ yr$^{-1}$ the 32313 expected O stars from the canonical IMF have to be multiplied by 0.51 and therefore only 16157 O stars are to be expected.}
\begin{tabular}{crrrrrrrr}
&B&O&B&O&B&O&B&O\\
&3-18 $M_\odot$&18$^+$ $M_\odot$&3-18 $M_\odot$&18$^+$
$M_\odot$&3-18 $M_\odot$&18$^+$ $M_\odot$&3-18 $M_\odot$&18$^+$ $M_\odot$\\
&&&$\alpha_3$ = 2.35&$\alpha_3$ = 2.35&$\alpha_3$ = fn($\rho_\mathrm{ecl}$)&$\alpha_3$ = fn($\rho_\mathrm{ecl}$)&$\alpha_3$ = fn($\rho_\mathrm{ecl}$)&$\alpha_3$ = fn($\rho_\mathrm{ecl}$)\\
SFR&&&$\beta$ = 2&$\beta$ = 2&$\beta$ = 2&$\beta$ = 2&$\beta$ = fn(SFR)&$\beta$ = fn(SFR)\\
$[M_{\odot} {\rm yr}^{-1}]$&IMF&IMF&IGIMF I&IGIMF I&IGIMF II&IGIMF II&IGIMF III&IGIMF III\\
1&2&3&4&5&6&7&8&9\\
\hline
$10^{-5}$&                     4&                 0  &0&0&0&0&0&0\\
$10^{-4}$&                   36&              3    &0.55&0&       0.55&            0&                0.55&        0\\
$10^{-3}$&                 356&            31   &0.66&0.04&  0.66&           0.04&           0.66&           0.04\\
$10^{-2}$&              3559&           313   &0.77&0.25& 0.77&            0.25&          0.77&            0.25\\
$10^{-1}$&            35594&         3135  &0.83&0.40& 0.83&             0.40&          0.83&             0.40\\
$10^{0}$ &          355939&       31347  &0.86&0.51& {\bf 0.87}&{\bf 0.58}&{\bf 0.87}&{\bf 0.55}\\
$10^{1}$ &        3559386&     313472  &0.88&0.58& {\bf 0.90}&{\bf 0.88}&{\bf 1.06}&{\bf 1.31}\\
$10^{2}$ &      35593862&   3134724 &0.90&0.64& {\bf 0.89}&{\bf 1.22}&{\bf 1.02}&{\bf 2.17}\\
$10^{3}$ &   355938622&  31347240 &0.91&0.68& {\bf 0.84}&{\bf 1.54}&{\bf 0.92}&{\bf 2.57}\\
$10^{4}$ & 3559386217&313472403&0.92&0.71& {\bf 0.79}&{\bf 1.80}&{\bf 0.82}&{\bf 2.84}\\
\hline
\end{tabular}
\end{table*}

\begin{figure*}
\begin{center}
\includegraphics[width=8cm]{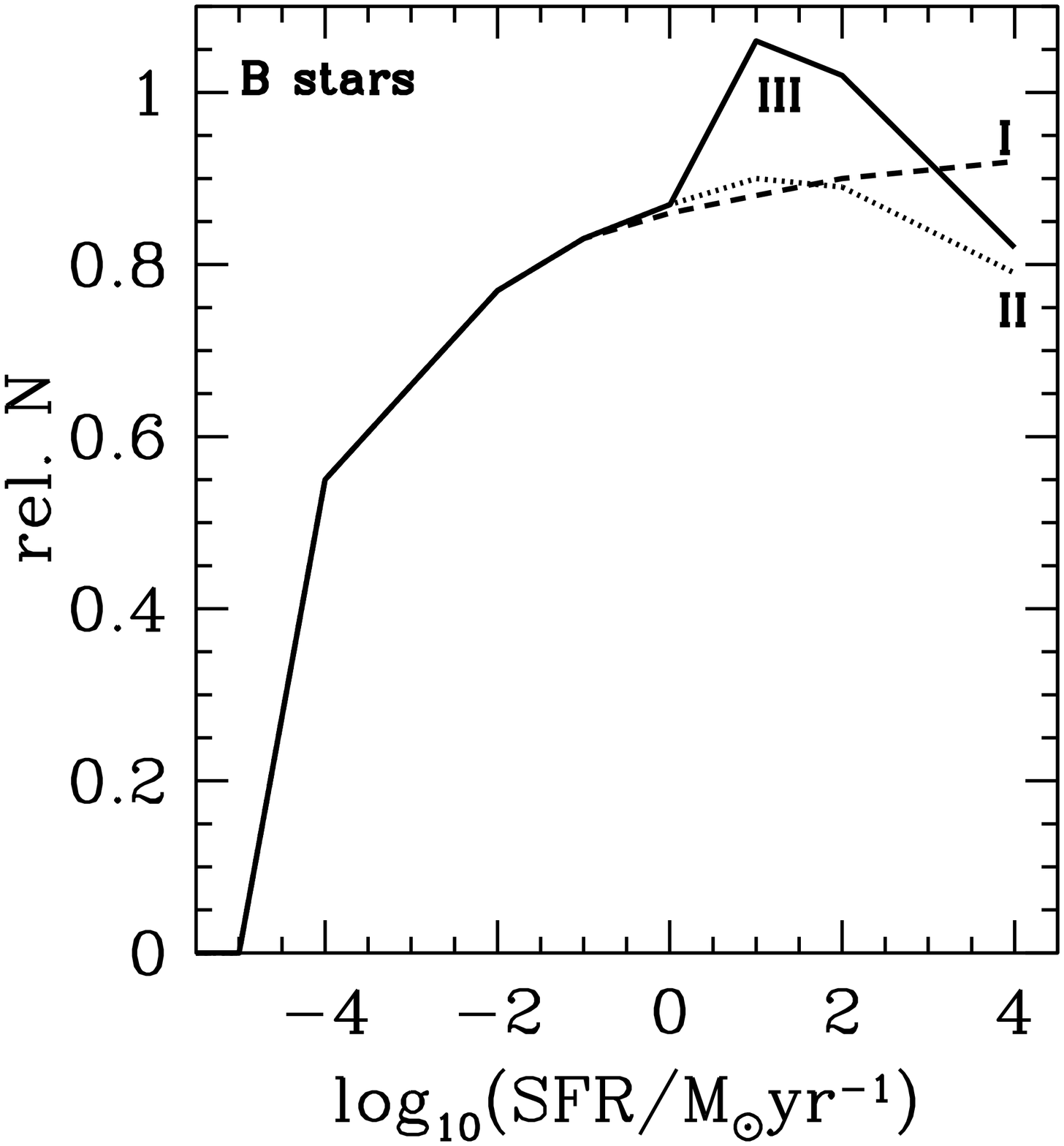}
\includegraphics[width=8cm]{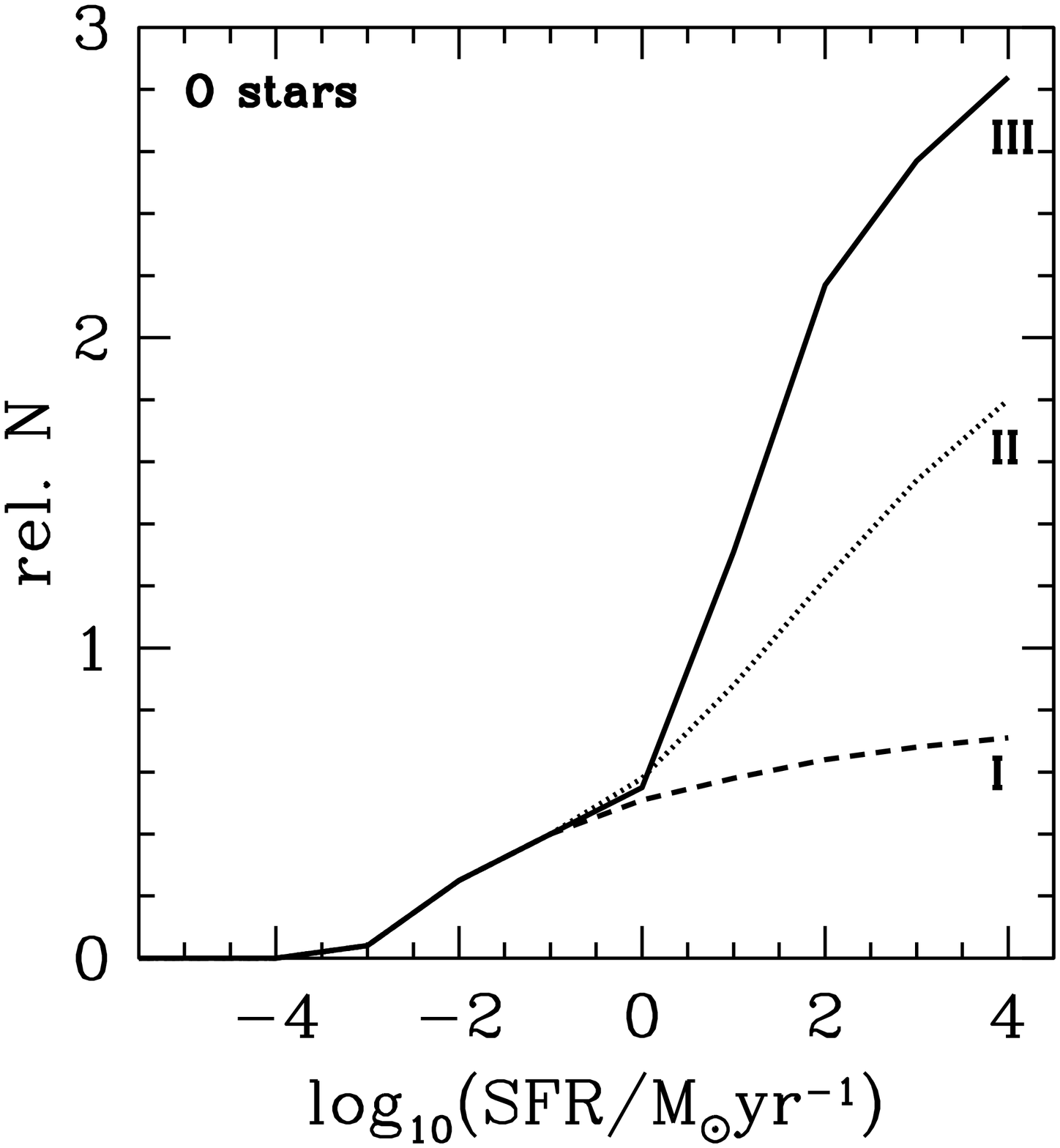}
\vspace*{-1.5cm}
\caption{The number of O and B stars in dependence of the SFR relative to an invariant canonical IMF for IGIMF model I (dashed lines), IGIMF model II (dotted lines) and IGIMF model III (solid lines). See also Table~\ref{tab:numbers}. For example, at a SFR = 1000 $M_\odot$ yr$^{-1}$ there ought to be 1.6 times as many O stars in the IGIMF model II than if the IGIMF were identical to the canonical IMF. The O stars correspond to objects with an absolute $B$ magnitude of $-$4 (for a 18 $M_\odot$ main-sequence star of solar metallicity) and below while B main-sequence stars would have $M_B$ = 0.46 to $-$4.}
\label{fig:3}
\end{center}
\end{figure*}

Table~\ref{tab:numbers} is visualised in Fig.~\ref{fig:3}. It shows that B stars are generally not very well suited to differentiate between the models. For SFRs above $10^{-3}$ $M_\odot$ yr$^{-1}$, the number of B stars is always roughly within 20\% of the value for an invariant IMF. Only for SFRs below $10^{-3}$ $M_\odot$ yr$^{-1}$, does the number of B stars in the IGIMF models drop significantly below what would be expected for the canonical IMF. In the case of O stars the situation is quite different. For SFRs below $10^{-1}$ $M_\odot$ yr$^{-1}$, the expected number of O stars for the IGIMF models are always much below the canonical IMF values. And for high SFRs (above $1$ $M_\odot$ yr$^{-1}$) it becomes possible to separate the different IGIMF models. With 4\% to 71\% of the number of O stars, the values for the non-top-heavy IGIMF (IGIMF model I) are always lower than the canonical IMF predictions. The top-heavy IGIMF with constant $\beta$ (IGIMF model II) has 22\% to 80\% more O stars than the canonical IMF for SFRs $\ge 10^{2}$ $M_\odot$ yr$^{-1}$. In the case of the top-heavy IGIMF with variable $\beta$ for SFR $\ge 1$ $M_\odot$ yr$^{-1}$ (IGIMF model III), the number of O stars is 31\% to 184\% larger than would be expected from the invariant canonical IMF.

\begin{figure}
\begin{center}
\includegraphics[width=8cm]{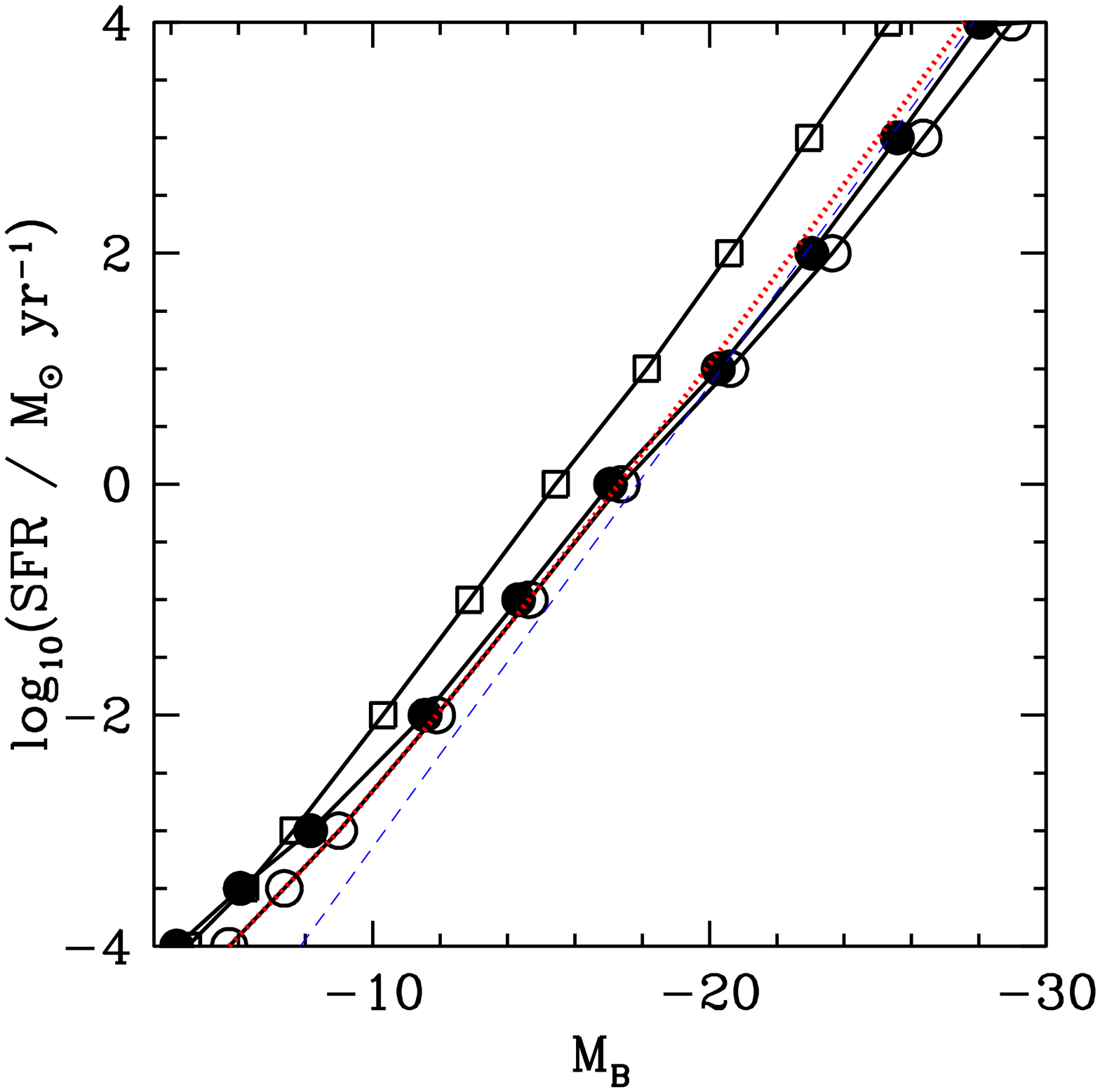}
\vspace*{-1.5cm}
\caption{The true total SFR as a function of absolute B band magnitude of the integrated population. The solid line with open circles is IGIMF model III with an assumed age of 0 Myr for the population. The solid line with filled dots is IGIMF model III but for an age of the population of 10 Myr, while the solid line with boxes marks IGIMF model III at 50 Myr. The (red) dotted line is IGIMF model I at 0  Myr and the thin (blue) dashed line is a model using the canonical IMF (Appendix~\ref{app:IMF}).}
\label{fig:Bmag}
\end{center}
\end{figure}

A different way to visualise these results is, for example, to plot how the $B$ band magnitude, M$_\mathrm{B}$, changes with SFR for the integrated population. This is done in Fig.~\ref{fig:Bmag}, though the impact of the IGIMF on optically visible photometric bands is unfortunately rather small.

\citet{HA10} studied the expected number of Milky Way O stars detectable by the {\it GAIA} astrometry mission and predict that for a SFR of 1 $M_\odot$ yr$^{-1}$ between 1000 and 4000 O stars should be observable for IGIMF models with different assumptions on the underlying IMF, the cluster mass function and the sampling method. As can be seen in Table~\ref{tab:numbers} the predicted number of O stars for a SFR of 1 $M_\odot$ yr$^{-1}$ is between 16157 and 40400. The factor of about 10 between both results originates from the assumption in \citet{HA10} that {\it GAIA} will only be able to observe about 10\% of the Galactic O stars and therefore, both results agree.

\section{Discussion \& Conclusions}
\label{se:diss}

Whether or not the IMF of whole galaxies, the IGIMF, is identical to the IMF as observed in individual pc-scale star formation events or in star clusters is of great importance for the modelling of stellar populations, the chemical evolution of galaxies and our understanding of star-formation in general. While only little to no indications for systematic variations of the slope of the IMF in star clusters in the Milky Way and the Magellanic Clouds have been found so far \citep{BCM10,KWP13}, increasing evidence for a variable galaxy-wide IMF is seen in unresolved extragalactic observational data \citep{LGB05,VD07,Da07,E08,WHT08,HAJ10,DKP12}. Multi-wavelength and spectroscopic studies of large volume-limited samples of galaxies give further evidence for systematic IMF variations. Both late-type \citep{HG07,MWK09,LGT09,GHS10} and early-type  \citep{VCP96,VPB97,CMA12,CSA13,CMA13} galaxies show evidence for the same behaviour that at high SFRs the stellar populations in galaxies are top-heavy. Furthermore, evidence for a variation of the IMF is also available for resolved populations as \citet{Sc86} found a steeper than canonical IMF for the MW field and \citet{UMM07} found a similar result for the nearby dwarf starburst galaxy NGC 4214 which has a SFR of about 0.2 $M_\odot$ yr$^{-1}$. Recently, a top-light IMF has also been found to have been active in the Fornax dwarf galaxy \citep{LCZ13} and the Sagittarius dwarf galaxy \citep{MWM13} precursors, in good agreement with the IGIMF theory \citep{KWP13}.

Adopting the observed correlations and distribution functions in star-forming galaxies (the 7 axioms of the IGIMF theory presented in \S~\ref{sub:axioms} and \S~\ref{sub:late}), it has become evident that the IMF of a whole galaxy must differ from the canonical IMF, thus implying the IGIMF to vary with SFR. 

One possible explanation for the top-heavy IMFs at very high star-formation rates could be due to cosmic-ray heating of Giant Molecular Clouds (GMCs) in starbursts \citep{Pa10,PTM11,PT13} which may have a two-fold impact on the IGIMF. It might prefer more massive clouds to collapse, thus leading to an increase of $M_\mathrm{ecl,min}$ and/or a decrease of $\beta$, but also within the most-massive clusters it can induce a top-heavy IMF as discussed in \citet{WK07a} and \citet{MKD10} by inhibiting fragmentation of molecular cloud cores. We correct eq.~14 of \citet{MKD10} (our eq.~\ref{eq:beta}). In dwarf galaxies on the other hand, the lack of shear forces might prefer the collapse of GMCs into a single or very few clusters instead of many \citep{WBZ10}, thus changing $M_\mathrm{ecl,min}$ for such objects as well.

With the \citet{GHS10} results on star-formation in late-type galaxies it is possible to deduce a relation between the power-law index of the ECMF, $\beta$, and the SFR (eq.~\ref{eq:beta}), suggesting a reduction of the formation of low-mass star clusters in starbursts (axiom 7 of the IGIMF theory). That is, we have un-earthed that the embedded cluster mass function may become top-heavy in starbursting galaxies. Remarkably, with this result it thus follows that the observed change of the M/L values of E galaxies with their mass can be readily understood within the IGIMF framework without further adjustment.

While the results presented with this contribution demonstrate that the IGIMF theory is in good agreement with the latest observational data it is necessary to further test it. To facilitate such a test the expected numbers of O and B stars in galaxies with different SFR are presented. They can be used to directly compare observational results with theoretical expectations. The best regime to discriminate between the IGIMF and the canonical IMF is to look for O stars in dwarf galaxies with SFRs around $10^{-2}\,M_\odot$ yr$^{-1}$. Here only around 80 O stars are expected when the canonical IMF predicts about 320. Likewise, lower SFRs are suitable, too. The different IGIMF models evaluated here are indistinguishable from each other at low SFRs.  For high SFRs the IGIMF models I--III studied here predict different very large numbers of O stars but these would be far too many for direct counting. Only integrated quantities like the H$\alpha$ luminosity can be used for observational tests in these cases. But note that the here used SFRs are the total SFRs, rather than H$\alpha$-based SFRs which sensitively depend on the number of O stars. B stars are only suitable to distinguish between an IGIMF and the canonical IMF for SFRs below $10^{-4}\,M_\odot$ yr$^{-1}$. Above this limit their numbers only deviate mildly from the values for the canonical IMF.

While still widely used in cosmology and extragalactic stellar population studies, an invariant IMF for galaxies is  excluded by mounting observational evidence. These independent observational results are readily explained by the IGIMF theory and corresponding models fit relatively well (Figs.~\ref{fig:GAMA2} and \ref{fig:m2l}), showing that the IGIMF is a relevant description of stellar populations in galaxies. This conclusion is independent of whether a constant SFH or an exponentially declining one is used. The IGIMF theory, based on the knowledge of the local star-formation process, is therefore a useful description of large-scale star-formation in whole galaxies. Thus with our knowledge of star-formation in the Milky Way, it is in principle possible to explain all observed IMF variations in extragalactic sources.

An extension of this work to include the results of \citet{FBR13}, who found a strong dependence of the galaxy-wide IMF power-law index on the central velocity dispersion of elliptical galaxies, is currently underway.


\section*{Acknowledgements}
We thank Prof.~G.~Meurer for helpful suggestions and comments. This work has been supported by the Programa Nacional de Astronom{\'i}a y Astrof{\'i}sica of the Spanish Ministry of Science and Innovation under grant AYA2010-21322-C03-02.


\begin{appendix}
\section{The canonical IMF}
\label{app:IMF}
The following two-component power-law stellar IMF is used throughout the paper:

{\small
\begin{equation}
\xi(m) = k \left\{\begin{array}{ll}
k^{'}\left(\frac{m}{m_{\rm H}} \right)^{-\alpha_{0}}&\hspace{-0.25cm},m_{\rm
  low} \le m < m_{\rm H},\\
\left(\frac{m}{m_{\rm H}} \right)^{-\alpha_{1}}&\hspace{-0.25cm},m_{\rm
  H} \le m < m_{0},\\
\left(\frac{m_{0}}{m_{\rm H}} \right)^{-\alpha_{1}}
  \left(\frac{m}{m_{0}} \right)^{-\alpha_{2}}&\hspace{-0.25cm},m_{0}
  \le m < m_\mathrm{1},\\ 
 \left(\frac{m_{0}}{m_{\rm H}} \right)^{-\alpha_{1}}
 \left(\frac{m_{1}}{m_{0}} \right)^{-\alpha_{2}}
  \left(\frac{m}{m_\mathrm{1}} \right)^{-\alpha_{3}}&\hspace{-0.25cm},m_{0}
  \le m < m_\mathrm{max},\\ 
\end{array} \right. 
\label{eq:4pow}
\end{equation}
\noindent with exponents
\begin{equation}
          \begin{array}{l@{\quad\quad,\quad}l}
\alpha_0 = +0.30&m_\mathrm{low} = 0.01 \le m/{M}_\odot < m_\mathrm{H} = 0.08,\\
\alpha_1 = +1.30&0.08 \le m/{M}_\odot < 0.50,\\
\alpha_2 = +2.35&0.50 \le m/{M}_\odot \le 1.00,\\
\alpha_3 = +2.35&1.00 \le m/{M}_\odot \le m_\mathrm{max}.\\
          \end{array}
\label{eq:imf}
\end{equation}}
\noindent where $dN = \xi(m)\,dm$ is the number of stars in the mass interval $m$ to $m + dm$. The exponents $\alpha_{\rm i}$ represent the standard or canonical IMF \citep{Kr01,Kr02,KWP13}. For a numerically practical formulation see \citet{PAK06}. An equivalent log-normal form is provided by eq.~4-56 (fig.~4-28) in \citet{KWP13}.

The advantages of the multi-part power-law description are the easy integrability and, more importantly, that {\it different parts of the IMF can be changed readily without affecting other parts}. Note that this form is a two-part power-law in the stellar regime, and that brown dwarfs contribute about 1.5 per cent by mass only and that differing binary properties near $m_\mathrm{H}$ implies most brown dwarfs to be a separate population \citep[$k^{'} \approx \frac{1}{3}$,][]{TK07,TK08}.

The observed IMF is today understood to be an invariant Salpeter/Massey power-law slope \citep{Sal55,Mass03} above $0.5\,M_\odot$, being independent of the cluster density and metallicity for metallicities $Z \ge 0.002$ \citep{MH98,SND00,SND02,PaZa01,Mass98,Mass02,Mass03,WGH02,BMK03,PBK04,PAK06}.  Furthermore, un-resolved multiple stars in the young star clusters are not able to mask a significantly different slope for massive stars \citep{MA08,WK07c}. \citet{Kr02} has shown that there are no trends with present-day physical conditions and that the distribution of measured high-mass slopes, $\alpha_3$, is Gaussian about the Salpeter value thus allowing us to assume for now that the stellar IMF is invariant and universal in each pc-scale star-formation event. There is evidence of a maximal mass for stars \citep[$m_{\rm max*}\,\approx\,150\,M_{\odot}$,][]{WK04}, a result later confirmed by several independent studies \citep{OC05,Fi05,Ko06}. However, according to \citet{CSH10} $m_\mathrm{max*}$ may also be as high as 300 $M_\odot$; though \citet{BK12} could show that these super-massive objects are very likely mergers of star formed with 150 $M_\odot$ or less. \citet{MKD10} uncovered a systematic trend towards top-heaviness (small $\alpha_3$) with increasing star-forming cloud density (see eq.~\ref{eq:topheavy}).

\end{appendix}

\bibliography{mybiblio}

\bsp
\label{lastpage}
\end{document}